\documentclass[astra]{copernicus}

\begin{document}
\title{The IceCube neutrino observatory: Status and initial results}

\author{Timo Karg for the IceCube
  collaboration\footnote{http://icecube.wisc.edu/}}
\affil{Bergische Universit\"at Wuppertal, Fachbereich C ---
  Mathematik und Naturwissenschaften, 42097 Wuppertal, Germany}

\runningtitle{The IceCube neutrino observatory}
\runningauthor{T.~Karg}
\correspondence{T.~Karg (karg@uni-wuppertal.de)}

\received{}
\pubdiscuss{} 
\revised{}
\accepted{}
\published{}

\firstpage{1}
\maketitle

\begin{abstract}
  The IceCube collaboration is building a cubic kilometer scale
  neutrino telescope at a depth of $2 \unit{km}$ at the geographic
  South Pole, utilizing the clear Antarctic ice as a Cherenkov medium
  to detect cosmic neutrinos. The IceCube observatory is complemented
  by IceTop, a square kilometer air shower array on top of the in-ice
  detector. The construction of the detector is nearly finished with
  79 of a planned 86 strings and 73 of 80 IceTop stations
  deployed. Its completion is expected in the winter 2010/11. Using
  data from the partially built detector, we present initial results
  of searches for neutrinos from astrophysical sources such as
  supernova remnants, active galactic nuclei, and gamma ray bursts,
  for anisotropies in cosmic rays, and constraints on the dark matter
  scattering cross section. Further, we discuss future plans and R\&D
  activities towards new neutrino detection techniques.
\end{abstract}

\introduction

The origin and acceleration of cosmic rays to energies beyond $10^{20}
\unit{eV}$ is one of the big open questions in astroparticle physics
today. Astrophysical objects that are promising source candidates
include supernova remnants or microquasars in our own galaxy, or, for
cosmic rays at the highest energies, extragalactic objects like active
galactic nuclei or gamma-ray bursts. Carrying electric charge, the
cosmic rays are deflected in magnetic fields during their propagation
and possibly do not point back to their source; an effect which is
stronger at lower energies. However, the cosmic rays being hadrons,
interactions in matter or photon fields at the source should produce
high energy neutrinos that would arrive at Earth undeflected and their
energy spectrum will carry valuable information of the physics
processes at the source. For a recent review see
e.g.~\citet{Anchordoqui:2010fk}.

To detect the small expected flux of astrophysical neutrinos large
detector volumes are necessary. Currently, the most competitive limits
on the neutrino flux in the TeV and PeV energy range are placed by
Cherenkov telescopes using natural transparent media like water or ice
as a detection medium. These include fresh water, e.g.~the Baikal
neutrino telescope \citep{Aynutdinov:2009kx} in Lake Baikal, sea
water, e.g.~the ANTARES neutrino telescope \citep{Coyle:2009fk} in the
Mediterranean Sea, or glacial ice as used for the IceCube neutrino
observatory presented here.

\section{The IceCube observatory}

The IceCube observatory is currently under construction at the
geographic South Pole. It comprises a $1 \unit{km}^3$ in-ice detector
measuring Cherenkov light from charged particles traversing the ice
and the IceTop air shower array on the surface. The in-ice detector
consists of $5160$ digital optical modules (DOMs) installed on $86$
vertical strings instrumented at depths between $1450 \unit{m}$ and
$2450 \unit{m}$ and deployed into the ice using a hot water
drill. Each DOM consists of a photomultiplier tube (PMT)
\citep{Abbasi:2010ys} housed in a glass pressure sphere and
electronics to digitize, timestamp, and transmit signals to the
central data acquisition system \citep{Abbasi:2009qf} located in the
IceCube laboratory on the surface. IceTop is made up of $160$
clear-ice tanks, each equipped with a low-gain and a high-gain
DOM. Pairs of IceTop tanks, called stations, are installed below the
snow surface, on top of the in-ice strings. Recent results from IceTop
have been presented at this conference \citep{Feusels, Kislat}.

\begin{figure}[t]
  \vspace*{2mm}
  \begin{center}
    \includegraphics[width=8.3cm]{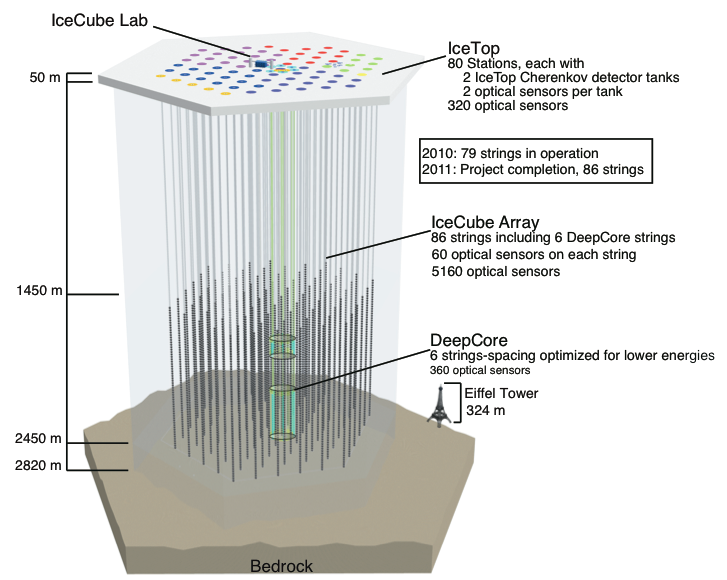}
  \end{center}
  \caption{Schematic of the IceCube neutrino observatory.}
  \label{fig:setup}
\end{figure}

The IceCube observatory with all its components is schematically shown
in Fig.~\ref{fig:setup}.  Eighty strings of the in-ice array are
installed on a regular triangular grid with a spacing of $125
\unit{m}$. In their center, in the deep, exceptionally clear ice, six
additional strings with high quantum efficiency DOMs and with closer
spacing, both horizontally and between the DOMs are installed, forming
the DeepCore array. Using the outer strings as a veto against
atmospheric muons, DeepCore is designed to lower the energy threshold
of IceCube to $< 100 \unit{GeV}$ \citep{Wiebusch:2009fk}. Currently,
in the austral winter 2010, $79$ of the $86$ IceCube strings,
including the six DeepCore strings, and $73$ out of $80$ IceTop
stations are deployed and successfully taking data. The construction
will be completed in the austral summer 2010/2011.

The trigger rate of the currently operating 79 string detector is
approximately $2.3 \unit{kHz}$, mostly caused by muons from extensive
air showers penetrating the ice into the deep in-ice
detector. Atmospheric neutrinos only contribute about $30 \unit{mHz}$
to the overall trigger rate. Although atmospheric muons are a valuable
tool for calibration, discriminating neutrinos from the atmospheric
background is the big challenge in the experiment. It is achieved by
using the Earth as a filter against atmospheric muons and searching
for events with an upward-going track as a signature, originating from
muons produced in neutrino interactions in the ice or bedrock around
the detector. Due to the neutrino-nucleon cross section increasing
with increasing neutrino energy, the Earth becomes opaque to neutrinos
at PeV energies. At these energies, the characteristic neutrino
signature are downward-going tracks, and neutrinos are discriminated
from atmospheric muons by the harder energy spectrum expected from
cosmic sources. An astrophysical neutrino flux would eventually emerge
as a hard component from the measured energy spectrum.

Apart from muon neutrinos, identified by the ``track like'' signature
of the muon produced in a charged current interaction, IceCube is also
sensitive to all other neutrino flavors. Electromagnetic and/or
hadronic cascades developing in interactions of neutrinos of all
flavors inside the detector volume are detected as ``point like''
sources of light due to the large spacing of the DOMs compared to the
dimensions of the cascade. Here, the signature of a neutrino is a
cascade observed in the detector without an incoming particle track.

\section{Recent results}

IceCube has delivered a variety of interesting and competitive
scientific results already during its construction phase, taking data
with the partially completed detector. In the following, selected
recent results from the IceCube neutrino observatory will be
presented.

\subsection{Neutrino point sources}

The IceCube collaboration has performed several different searches for
point-like neutrino sources, including all sky searches for steady
sources, sources variable in time, and observing pre-selected sources
of special interest.

\textit{Time integrated search.} For the first time, an all sky search
has been performed with the data from the IceCube 40 string
configuration measured during $375.5$ days of live time in 2008 and
2009. The analyzed data consist of approximately $40\,000$ track like
events. $40\%$ of the events originate from the northern hemisphere
(upward-going events) and are dominated by atmospheric neutrinos in
the ten to a few hundred TeV energy range. The remaining events,
coming from the southern hemisphere, pre-dominantly consist of high
energy atmospheric muons propagating into the in-ice detector. They
are selected to have typical energies in the PeV range where the flux
of cosmic neutrinos is expected to emerge from the softer atmospheric
muon spectrum.

\begin{figure*}[t]
  \vspace*{2mm}
  \begin{center}
    \includegraphics[width=0.67\textwidth]{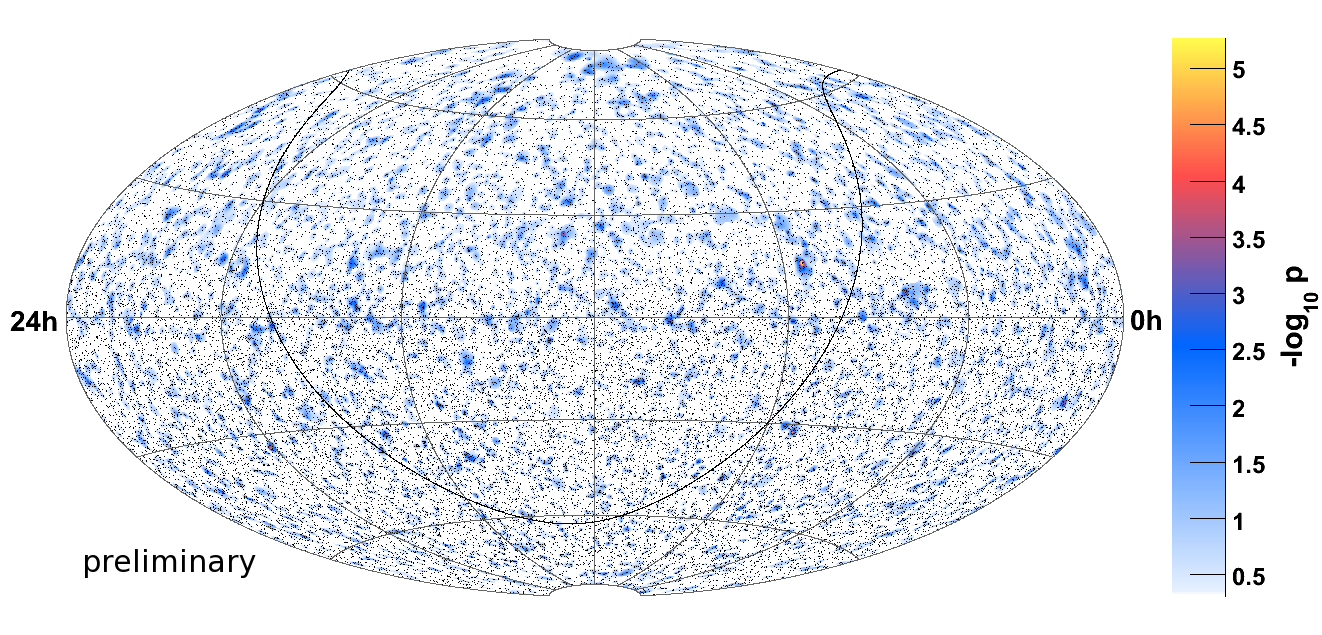}
  \end{center}
  \caption{IceCube 40 string sky map and pre-trial $p$-values. The
    black dots represent the directions of all events used in the
    analysis. The color scale represents the pre-trial $p$-values.}
  \label{fig:IC40-ps}
\end{figure*}

The data are analysed using an unbinned likelihood ratio method
\citep{Braun:2008jl}. Based on the reconstructed direction of the
track and an energy estimator we search for an excess of events
exceeding the background hypothesis. The data are modeled as a two
component mixture of signal and background, leaving the source
strength and spectral slope as free parameters in the likelihood
maximization.  The sky map of all events used in the search is shown
in Fig.~\ref{fig:IC40-ps} together with the $p$-values calculated for
each direction. The highest significance (pre-trial $p$-value: $5.2
\cdot 10^{-6}$) is observed in the direction $113.75\degree$ right
ascension, $15.15\degree$ declination. In trials using scrambled data
sets, $18\%$ of all scrambled data sets have an equal or higher
significance somewhere in the sky.

\begin{figure}[t]
  \vspace*{2mm}
  \begin{center}
    \includegraphics[width=8.3cm]{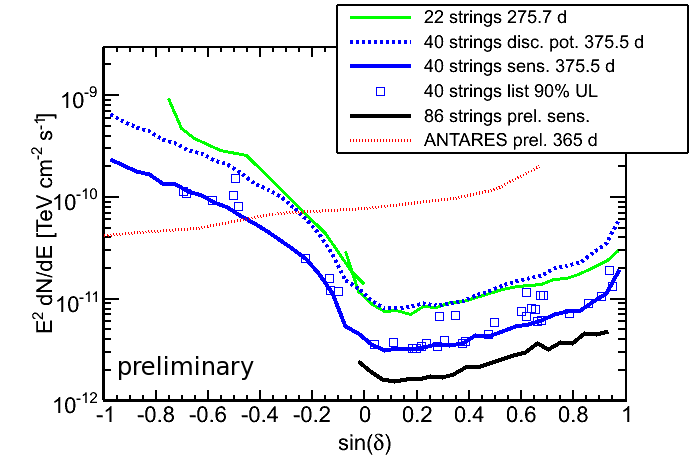}
  \end{center}
  \caption{Sensitivity for a point-like source with an $E^{-2}$ flux
    of muon neutrinos and anti-neutrinos as a function of
    declination. Upper limits ($90\%$ C.L.) on the flux from $35$ a
    priori selected sources are given (blue squares). For comparison,
    sensitivities from a previous analysis \citep{Abbasi:2009dd} and
    the predicted sensitivity for the full IceCube detector and the
    ANTARES experiment \citep{Coyle:2009fk} are indicated. In addition
    the discovery potential for the IceCube 40 string configuration is
    shown.}
  \label{fig:IC40-ps-limits}
\end{figure}

The non-observation of a neutrino point source allows us to place
upper limits on the neutrino flux from point-like sources. In
Fig.~\ref{fig:IC40-ps-limits} the sensitivity and discovery potential
for sources with an $E^{-2}$ energy spectrum are shown as a function
of declination. Upper limits on the muon (anti-)neutrino flux for $35$
a priori selected point-source candidates are indicated.

\begin{figure}[t]
  \vspace*{2mm}
  \begin{center}
    \includegraphics[width=8.3cm]{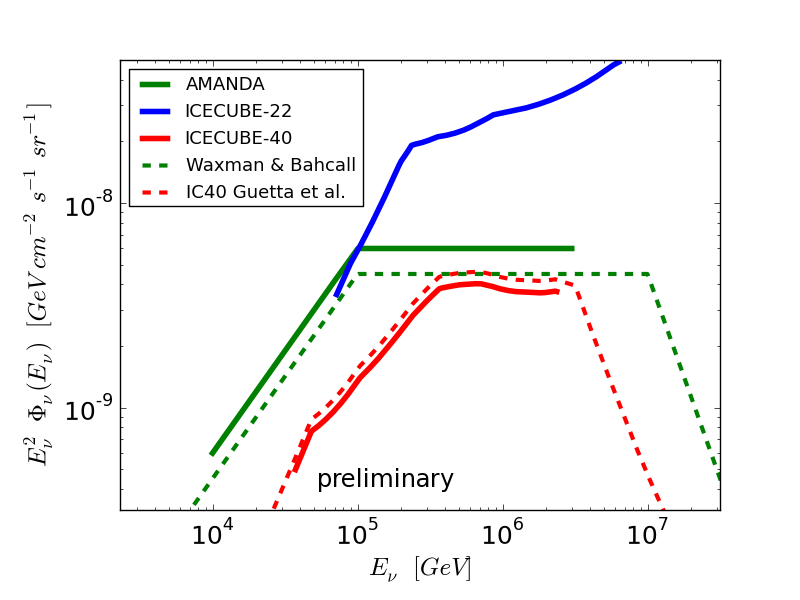}
  \end{center}
  \caption{IceCube 40 string upper limit (red curve) on the prompt
    neutrino flux from gamma-ray bursts ($90\%$ C.L.) compared to
    previous limits obtained with the AMANDA-II detector
    \citep{Achterberg:2008yq} and the IceCube 22 string configuration
    \citep{Abbasi:2010ly}. The dashed red curve is the expected flux
    summed over all $117$ individually modelled GRBs (see text for
    details).}
  \label{fig:IC40-grb}
\end{figure}

\textit{Neutrinos from gamma-ray bursts.} Gamma-ray bursts (GRBs),
like all transient astrophysical sources that are expected to produce
neutrinos, allow for very sensitive analyses since the expected
background can be largely reduced by requiring coincidence in both
direction and time with the observed event. We have performed a search
for prompt neutrino emission from $117$ satellite detected gamma-ray
bursts in the northern hemisphere in IceCube 40 string configuration
data. The expected neutrino flux from each GRB was individually
modeled using the model described in \citet{Guetta:2004fk}. We use an
unbinned maximum likelihood analysis, assigning each IceCube event a
signal probabiblity based on its angular and temporal distance from
the GRB and taking into account an energy estimator. No coincident
events have been observed and an upper limit on the prompt neutrino
flux from GRBs has been calculated. Fig.~\ref{fig:IC40-grb} shows the
model dependent upper limits at $90\%$ confidence level.

\textit{Further point source searches.} Several other searches for
neutrinos from point like sources have been performed with IceCube
data. Results from a stacking analysis using starburst galaxies
\citep{Dreyer} and from a search for spatial coincidence with the
highest energy cosmic rays observed by the Pierre Auger and HiRES
experiments \citep{Lauer} have been reported at this conference.

Further analyses on a-priori defined lists of point-source candidates,
including both, galactic and extra-galactic, and steady and
time-variable sources, have resulted in the most stringent upper
limits on the neutrino fluxes existing today for various source
classes.

\subsection{Diffuse neutrino fluxes}

The diffuse neutrino flux is constituted by atmospheric neutrinos,
cosmogenic neutrinos produced in the interaction of ultra high energy
cosmic rays with the cosmic microwave background radiation, and a
superposition of fluxes from unresolved point-like sources.

\begin{figure*}[t]
  \vspace*{2mm}
  \begin{center}
    \includegraphics[width=0.67\textwidth]{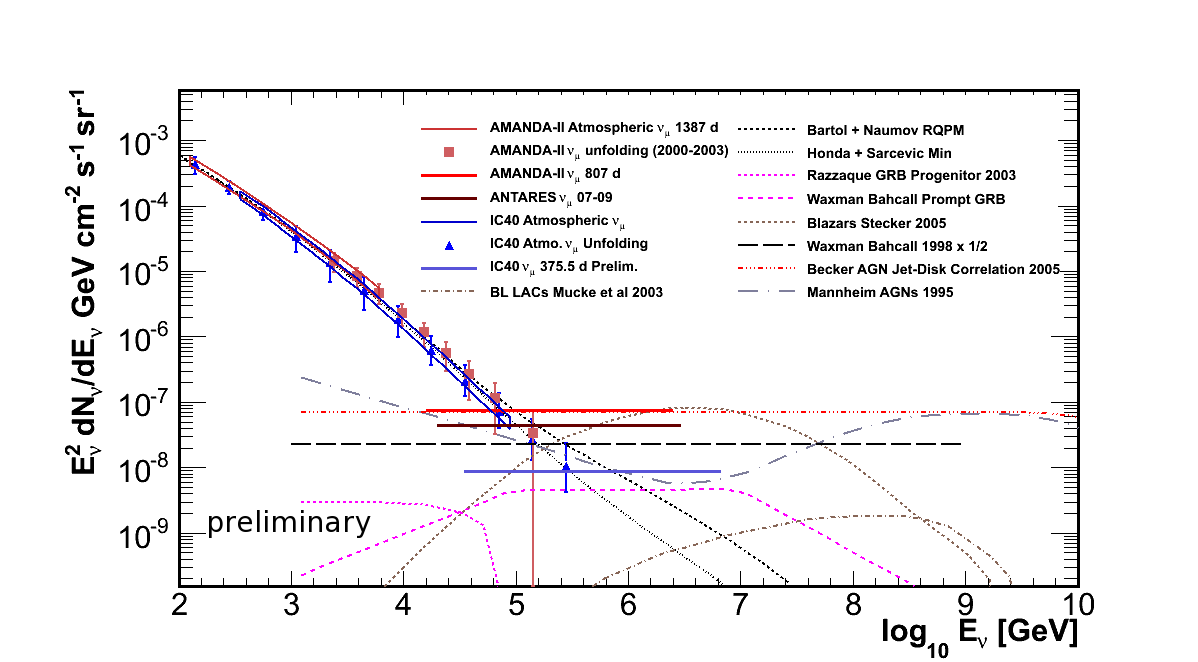}
  \end{center}
  \caption{Atmospheric muon neutrino spectrum measured with the
    IceCube 40 string configuration and upper limit ($90\%$ C.L.) on
    an $E^{-2}$ astrophysical diffuse neutrino flux (blue curves; see
    also \citet{Grullon:2010fk}). The blue triangles are the result
    from an independent unfolding analysis of IceCube 40 string data
    \citep{Abbasi:2010dq}. For comparison several experimental limits
    \citep{Abbasi:2009yq, Abbasi:2010uq, Achterberg:2007rt, Anton} and
    theoretical flux predictions for atmospheric \citep{Barr:2004fk,
      Enberg:2008vn, Fiorentini:2001uq, Honda:2007kx} and
    astrophysical \citep{Becker:2005fk, Mannheim:1995uq, Mucke:2003ys,
      Razzaque:2003zr, Stecker:2005kx, Waxman:1997dq, Waxman:1998vn}
    neutrinos are given.}
  \label{fig:IC40-diffuse}
\end{figure*}

We have performed a search for an excess of muon \hbox{(anti-)}
neutrinos on the atmospheric neutrino spectrum using the data from the
IceCube 40 string configuration. The data were modeled as a
composition of atmospheric neutrinos, prompt atmospheric neutrinos (a
hard component originating mainly from the decay of charm particles)
and an astrophysical contribution with an $E^{-2}$ energy
spectrum. The flux normalisation and corrections to the spectral shape
were left as free parameters in the likelihood function used. The
resulting atmospheric neutrino energy spectrum is in very good
agreement with model predictions. No excess above the expected flux of
atmospheric neutrinos has been observed. We place the currently most
stringent upper limits on a diffuse flux of astrophysical
neutrinos. At $90\%$ confidence level the upper limit on a muon
neutrino flux following an $E^{-2}$ spectrum is $E^2 \Phi < 8.9 \cdot
10^{-9} \unit{GeV} \unit{cm}^{-2} \unit{s}^{-1} \unit{sr}^{-1}$ in the
energy range $4.54 < \log_{10} (E_{\nu} / \unit{GeV}) <
6.84$. Fig.~\ref{fig:IC40-diffuse} shows the measured muon neutrino
energy spectrum and the upper limit on an $E^{-2}$ neutrino flux. We
exclude several of the more optimistic astrophysical neutrino
production models at the $5 \sigma$ level: the AGN jet-disk
correlation model \citep{Becker:2005fk}, the AGN jet model
\citep{Mannheim:1995uq}, and the blazar model
\citep{Stecker:2005kx}. We exclude a diffuse neutrino flux at the
level of the WB upper bound \citep{Waxman:1998vn} at the $3 \sigma$
level.

\subsection{Indirect dark matter searches}

IceCube also performs indirect searches for dark matter particles
gravitationally trapped in the center of the Sun and the Earth. Their
signature is a flux of neutrinos from a well defined direction
produced in the annihilation of dark matter particles into standard
matter, where the energy of the produced neutrinos depends on the mass
of the dark matter particles and the dominant annihilation channel.
The non-observation of an excess of neutrinos from the direction of
the center of the Sun, the Earth, or the galactic center allows us to
place upper limits on the dark matter-nucleon interaction cross
section as a function of the mass of the dark matter particle for
different dark matter models.

One promising dark matter candidate is the lightest supersymmetric
particle, the neutralino.  We have searched for an excess of neutrinos
from the direction of the Sun in the data of the IceCube 22 string
detector. No events above the expected background have been
observed. For neutralinos with spin-independent interactions with
ordinary matter, IceCube is only competitive with direct detection
experiments if the neutralino mass is sufficiently large. On the other
hand, for neutralinos with mostly spin-dependent interactions, IceCube
places the most stringent limits for neutralino masses above $250
\unit{GeV}$ \citep{Abbasi:2009vn}.

\subsection{Cosmic ray anisotropy}

With over $10^9$ cosmic ray induced atmospheric muons measured every
year in the in-ice detector, IceCube is well suited to study
anisotropies in the cosmic ray flux. We have analysed the arrival
directions of $4.3 \cdot 10^9$ atmospheric muons detected with the
IceCube 22 string configuration. The median energy of the cosmic ray
particles inducing the air showers is $20 \unit{TeV}$ and the muons
are reconstructed with a median angular resolution of $3
\unit{\degree}$. We observe a large scale anisotropy in the right
ascension of the arrival directions with a first harmonic amplitude of
$6.4 \cdot 10^{-4}$ \citep{Abbasi:2010vn}. Our result represents the
first measurement in the multi-TeV energy range covering the entire
southern hemisphere. The phase of the observed anisotropy matches the
one of previously measured cosmic ray anisotropies in the northern
hemisphere by the Tibet \citep{Amenomori:2006fk} and Milagro
\citep{Abdo:2009uq} experiments, indicating that we observe a
continuation of the effect measured by these experiments.

\section{R\&D activities}

To enhance the capabilities of the IceCube observatory the
collaboration conducts a vigorous R\&D program. The detection of the
small neutrino flux predicted at the highest energies ($E_\nu > 10^{8}
\unit{GeV}$) requires detector target masses of the order of $100$
gigatons, corresponding to $100 \unit{km}^3$ of ice. The optical
Cherenkov neutrino detection technique is not easily scalable from $1
\unit{km}^3$-scale telescopes to such large volumes. Promising
techniques with longer signal attenuation lengths, allowing for the
sparse instrumentation of large volumes of Antarctic ice, are the
radio and acoustic detection methods. The radio approach utilizes the
Askaryan effect, the coherent emission of radio waves from the charge
asymmetry developing in an electromagnetic cascade in a dense medium
\citep{Askaryan:1962fk}. Acoustic detection is based on the
thermo-acoustic sound emission from a particle cascade depositing its
energy in a very localized volume causing sudden expansion that
propagates as a shock wave perpendicular to the cascade
\citep{Askaryan:1979vn}.

The IceCube collaboration has developed and installed several test
setups to study the feasibility of these techniques at the South
Pole. This includes the development of sensors and data acquisition
systems suitable for Antarctic conditions and the measurement of the
properties of the ice relevant for the propagation and detection of
radio or acoustic signals. The quantities of interest include the
signal attenuation length, the noise level, the depth dependent index
of refraction (or sound speed accordingly), and the characterization
of possible transient backgrounds. IceCube's radio extension consists
of several radio frequency detectors for the frequency range from $100
\unit{MHz}$ up to $1 \unit{GHz}$ and calibration equipment deployed as
part of the IceCube array at depths between $5 \unit{m}$ and $1400
\unit{m}$ \citep{Landsman:2010}. The South Pole Acoustic Test Setup
(SPATS) comprises four strings instrumented with acoustic transmitters
and receivers co-deployed in IceCube boreholes at depths down to $500
\unit{m}$ \citep{Karg:2010}.

We also investigate the possibility of extending the IceTop air shower
detector with an array of radio antennas, measuring the coherent
geosynchrotron radiation emitted by air shower electrons and positrons
in the Earth's magnetic field. The radio signal in the frequency band
from a few MHz to $150 \unit{MHz}$ constitutes a third, complementary
measurement of the air shower properties in addition to the charged
particles on ground level measured with IceTop and the hard muon
component measured in the in-ice detector. We expect improved
sensitivity in the following three different physics analyses. Cosmic
ray composition studies will be enhanced through an independent
measurement of the depth of the shower maximum. Ultra high energy
cosmic ray photons can be detected over an increased range of zenith
angles as air showers without a muon component in the deep-ice. The
extended detection area can be used as a veto against air shower muon
bundles in the in-ice detector in searches for extremely high energy
neutrinos. As a first step towards a radio surface detector, a Radio
Air Shower Test Array (RASTA) is described in \citet{Boeser:2010}.

\conclusions

The IceCube observatory is very close to its completion and is
detecting neutrinos on a regular basis. About $20\,000$ neutrinos have
already been observed with the partially built detector. The measured
neutrino flux is still in good agreement with atmospheric
expectations, but we are beginning to explore astrophysically
interesting flux regions.

\begin{acknowledgements}
  We acknowledge the support from the following agencies:
  U.S.~National Science Foundation-Office of Polar Programs,
  U.S.~National Science Foundation-Physics Division, University of
  Wisconsin Alumni Research Foundation, U.S.~Department of Energy, and
  National Energy Research Scientific Computing Center, the Louisiana
  Optical Network Initiative (LONI) grid computing resources; National
  Science and Engineering Research Council of Canada; Swedish Research
  Council, Swedish Polar Research Secretariat, Swedish National
  Infrastructure for Computing (SNIC), and Knut and Alice Wallenberg
  Foundation, Sweden; German Ministry for Education and Research
  (BMBF), Deutsche Forschungsgemeinschaft (DFG), Research Department
  of Plasmas with Complex Interactions (Bochum), Germany; Fund for
  Scientific Research (FNRS-FWO), FWO Odysseus programme, Flanders
  Institute to encourage scientific and technological research in
  industry (IWT), Belgian Federal Science Policy Office (Belspo);
  Marsden Fund, New Zealand; Japan Society for Promotion of Science
  (JSPS); the Swiss National Science Foundation (SNSF), Switzerland;
  A.~Gro{\ss} acknowledges support by the EU Marie Curie OIF Program;
  J.~P.~Rodrigues acknowledges support by the Capes Foundation,
  Ministry of Education of Brazil.
\end{acknowledgements}

\bibliographystyle{copernicus}
\bibliography{karg.bib}
\end{document}